\documentclass{article}
\usepackage[conference]{IEEEtran}
\IEEEoverridecommandlockouts

\usepackage{cite}
\usepackage{amsmath,amssymb,amsfonts}
\usepackage{textcomp}
\usepackage{flushend}
\usepackage{cuted}
\usepackage{setspace}
\usepackage{multirow}
\usepackage{color}
\usepackage[table,xcdraw,dvipsnames]{xcolor}
\usepackage{xspace}
\usepackage{upquote}
\usepackage{url, listings, float}
\usepackage{graphicx, subfigure}
\usepackage[utf8]{inputenc}
\usepackage{enumitem}
\usepackage{answers}
\usepackage{tabularx}
\usepackage{threeparttable}
\usepackage{dblfloatfix}    
\usepackage{verbatim}
\usepackage{ifthen}
\usepackage{fancyhdr} 
\usepackage{array, boldline, makecell, booktabs}
\usepackage{tcolorbox}

\def\ie{\textit{i.e.,}~}
\def\etal{\textit{et al.}\xspace}

\def\eg{\textit{e.g.,}~}

\def\BibTeX{{\rm B\kern-.05em{\sc i\kern-.025em b}\kern-.08em
    T\kern-.1667em\lower.7ex\hbox{E}\kern-.125emX}}

\newboolean{showcomments}
\setboolean{showcomments}{true}
\newcommand{\todo}[1]{\ifthenelse{\boolean{showcomments}}
	{ \textcolor{green}{(To do:  #1)}}{}}
\newcommand{\note}[1]{\ifthenelse{\boolean{showcomments}}
	{ \textcolor{green}{(Cmt:  #1)}}{}}
\newcommand{\toall}[1]{\ifthenelse{\boolean{showcomments}}
    {\textcolor{red}{To All: #1}}{}}

\colorlet{punct}{red!60!black}
\definecolor{background}{HTML}{EEEEEE}
\definecolor{delim}{RGB}{20,105,176}
\colorlet{numb}{magenta!60!black}

\begin{document}

\title{Impact of COVID-19 on City-Scale Transportation and Safety: An Early Experience from Detroit}

\author{\IEEEauthorblockN{Yongtao Yao}
\IEEEauthorblockA{\textit{Department of Computer Science} \\
\textit{Wayne State University}\\
Detroit, MI 48202, USA \\
yongtaoyao@wayne.edu}
\and
\IEEEauthorblockN{Tony G. Geara}
\IEEEauthorblockA{\textit{Department of Public Works} \\
\textit{City of Detroit}\\
Detroit, MI 48216, USA \\
gearat@detroitmi.gov}
\and
\IEEEauthorblockN{Weisong Shi}
\IEEEauthorblockA{\textit{Department of Computer Science} \\
\textit{Wayne State University}\\
Detroit, MI 48202, USA \\
weisong@wayne.edu}
}

\maketitle


\begin{abstract}
The COVID-19 pandemic brought unprecedented levels of disruption to the local and regional transportation networks throughout the United States, especially the Motor City---Detroit. That was mainly a result of swift restrictive measures such as statewide quarantine and lock-down orders to confine the spread of the virus and flatten-the-curve along with a natural reaction of the population to the rising number of COVID-19-related cases and deaths. This work is driven by analyzing five types of real-world data sets from Detroit related to: traffic volume, daily cases, weather, social distancing index, and crashes from January 2019 to June 2020. The primary goal is figuring out the impacts of COVID-19 on the transportation network usage (traffic volume) and safety (crashes) for the City of Detroit, exploring the potential correlation between these diverse data features, and determining whether each type of data (\eg traffic volume data) could be a useful factor in the confirmed-cases prediction. In addition, early future prediction of COVID-19 rates can be a vital contributor to live-saving advanced preventative and preparatory responses. In order to achieve this goal, a deep learning model was developed using long short-term memory networks to predict the number of confirmed cases within the next one week. The model demonstrated a promising prediction result with a coefficient of determination ($R^2$) of up to approximately 0.91. Moreover, in order to provide statistical evaluation measures of confirmed-case prediction and to quantify the prediction effectiveness of each type of data, the prediction results of six feature groups are presented and analyzed. Furthermore, six essential observations with supporting evidence and analyses are presented. Those will be helpful for decision-makers to take specific measures that aid in preventing the spread of COVID-19 and protect public health and safety. The goal of this paper is to present a proposed approach which can be applied, customised, adjusted, and replicated for analysis of the impact of COVID-19 on a transportation network and prediction of the anticipated COVID-19 cases using a similar data set obtained for other large cities in the USA or from around the world. 
\end{abstract}


\section{Introduction}
The 2019 Novel Coronavirus (SARS-CoV-2), commonly known as COVID-19, has spread rapidly across the globe. As of July 28, 2020, over 16 million confirmed cases and 650 thousand deaths had been reported worldwide (WHO Situation Report-190, 2020) \cite{world2020coronavirus}; meanwhile, the United States is one of the most affected nations in the world, with more than 4 million confirmed cases and 149 thousand deaths. This tragic spread of COVID-19 nationally has resulted in disparate impacts across states and cities.

To slow the progression of COVID-19 and limit fatalities, public officials throughout Michigan had published a series of government directives that have been changed over time, starting with voluntary requests for stay-at-home and restrictions on large public gatherings, and then, statewide quarantine and lock-down orders. Nonetheless, essential travel activities continue to take place across Detroit, such as people's access to daily supplies, medical services, and other basic necessities of welfare and safety. These government directives inevitably affect various forms of travel activities and then impact transportation across Detroit significantly \cite{hu2020impacts}. 

Our work is based on the hypothesis that the objective, reliable and continuous transportation data can reflect the degree of social distancing, \ie the possibility of social activities and interpersonal communication to a certain extent, while many previous works provide the evidence proving that the social distancing measures enacted have led to control of COVID-19 \cite{ainslie2020evidence, briscese2020compliance, courtemanche2020strong}. Therefore, we believe that traffic data can provide a basis for the current and incoming pandemic status, and it is meaningful to explore the changes in traffic patterns during the COVID-19 pandemic for a specific city. 

In addition, crash-related information, such as total number of daily crashes, severity, and crash type can indirectly reflect traffic conditions \cite{stutts2001role, candefjord2016prehospital, abdel2004predicting}. Since one of the focus points of this work is to explore the correlation between traffic volume data and the outbreak of COVID-19, we also collected and analyzed crash data from Detroit to identify the impacts of crash-related information on the confirmed-cases prediction. 

Through our literature review, an abundance of studies pointed out that weather factors, \eg temperature (°C) and wind speed (mph) can contribute to the spread of COVID-19 \cite{tosepu2020correlation, gupta2020effect, csahin2020impact}. Inspired by these works, we also sought to determine whether weather could be a factor in the spread of this disease. 

Moreover, it is well known that maintaining social distancing can prevent the spread of COVID-19 disease and contain the number of casualties \cite{singh2020age, mohler2020impact, painter2020political, olivera2020keep, chen2020social, lewnard2020scientific}, which is based on the assumption that the degree of social distancing is highly related to the spread speed of COVID-19. Therefore, we also collected social distancing related data for Wayne County and Michigan state, to explore and test the correlation between social distancing and the severity of COVID-19 disease.

By considering the aforementioned data features, we then aimed to build an effective deep learning model using long short-term memory networks (LSTM) to predict the number of confirmed cases in Detroit. In order to provide statistical evaluation measures to quantify the prediction effectiveness of each type of data on the confirmed-cases prediction results, \ie the performance of LSTM, we trained LSTM on six experiment groups with different features, then analyzed the prediction results of the six feature groups. 

Our observations and prediction model are intended to help decision-makers to concentrate suitable public health efforts and apply effective transportation management techniques to protect residents and improve safety for Detroiters. It must be noted that the the presented statistical analysis approaches and the proposed prediction model were used on the Detroit-based data set as an example and due to availability. The method could be applied, customised, adjusted, and replicated for analysis of the impact of COVID-19 on a transportation network and prediction of the anticipated COVID-19 cases using a similar data set obtained from other large cities from within the USA or from around the world. 

Particularly, this paper set out to answer and is driven by the following question:
\begin{itemize}
    \item [$i$)] What are the sudden and drastic changes in overall temporal traffic patterns resulting from the outbreak of COVID-19?
    \item [$ii)$] Did traffic decrease and then recover evenly across all measured signals during COVID-19 and as COVID-19 restrictions were being lifted?
    \item [$iii)$] What are the impacts of COVID-19 and the social distancing on the reasons behind crashes? 
    \item [$iv)$] Can we leverage the traffic count data, crash data, and other COVID-19 related information, such as the data on daily confirmed cases plus social distancing and combined with weather information, to predict the number of COVID-19 confirmed cases for the next one week?

\end{itemize}

The rest of the paper is organized as follows. We first review previous works and introduce the research gap in Sec.~\ref{sec:related}. Then, we elaborate on the data sets used for the experiments in Sec.~\ref{sec:data}. Sec.~\ref{sec:statistic} describes experimental investigations for the questions presented in $i$) to $iii$), and Sec.~\ref{sec:method} demonstrates the proposed algorithms to predict the number of COVID-19 confirmed cases as described in question $iv$). After presenting a discussion of the prediction results in Sec.~\ref{sec:results}, we conclude the work in Sec.~\ref{sec:conclusion}.

\section{Related Work}
\label{sec:related}
In this section, we review recently published works that focus on the COVID-19 confirmed-cases prediction or similar research triggered by the outbreak of COVID-19 in terms of transportation, weather, social distancing, and other aspects. To the best of our knowledge,  prior works do not consider the effects of \textit{traffic volume data} and \textit{crash-related information} on the research of COVID-19 forecasts.

\vspace{-0.15cm}
\subsection{Transportation and COVID-19}
The travel restrictions put in place to reduce the spread of COVID-19 resulted in a sharp reduction in traffic throughout the United States. Some recent works explored the changes in the transportation mode. For example, in the work of \cite{hu2020impacts}, Hu \etal studied transportation modes during and after the COVID-19 pandemic using basic laws of traffic and mathematical analysis to explore scenarios of increased car commuting. Lacus \etal \cite{iacus2020estimating}  analyzed data on air traffic worldwide with the scope of analyzing the impact of the travel ban on the aviation sector as well as after changes in
COVID-19 diagnostic criteria. Lau \etal \cite{lau2020positive} calculated the correlation of air traffic to the number of confirmed COVID-19 cases and determined the growth curves of cases  before and after lock-down.  Teixeira \etal \cite{teixeira2020link} presented clues on how bike-sharing can support the transition to a post-COVID-19 society.

\subsection{Weather and COVID-19}

Moreover, the weather factors including temperature (°C), humidity(\%), wind speed (mph) are regarded as the factors that triggered the spread of COVID-19 in recent works. For example, through the Spearman-rank correlation test, Tosepu \etal \cite{tosepu2020correlation} proved that among the minimum temperature, maximum temperature, average temperature, humidity, and rainfall, only the average temperature is significantly related to the COVID-19 pandemic in Jakarta Indonesia. Another experiment \cite{gupta2020effect} was conducted based on the daily new cases and weather information in 50 states in the United States, and clarify those weather parameters \ie temperature and absolute humidity will help classify the risky geographic areas in different countries. 
Besides, based on the Spearman's correlation coefficients, Mehmet \etal \cite{csahin2020impact} pointed out the highest correlations between wind speed (mph) with the outbreak of COVID-19. In addition, a few studies have claimed that warm weather can possibly slow down the global pandemic \cite{csahin2020impact} of COVID-19 by considering nine cities in Turkey. 
 
\subsection{Social Distancing and COVID-19} 
In addition, a portion of the related work studied the relationship between social distancing and COVID-19. For instance, Singh \etal \cite{singh2020age} focused on the age-structured impact of social distancing on the COVID-19 epidemic in India and presented a mathematical model of the spread
of infection in a population that structured by age and social contact between ages. Courtemanche \etal \cite{courtemanche2020strong} pointed out that there would have been ten times greater spread of COVID-19 without shelter-in-place orders. \cite{mohler2020impact} presented that social distancing and shelter in place has had some impact on crime and disorder, but only for a restricted collection of crime types and not consistently across places.

\subsection{Other Coronavirus-Related Research}
 To date, more and more researchers are aggressively shifting their focus to detect, predict, treat, and recover from COVID-19. Some of the prior works do not consider transportation, weather, or social distancing, but they presented promising results to address the coronavirus-related issue. For example,  Qin \etal
 \cite{qin2020prediction} proposed a novel model to predict the outbreak of COVID-19 in populations in affected areas based on the social media search indexes (SMSI) for dry cough, fever, chest distress, coronavirus, and pneumonia.  Lacus \etal \cite{iacus2020estimating} calculated the economic impact measured in terms of loss of GDP due to the aviation sector as well as the social impact due to job losses related to aviation and correlated sectors (tourism, catering, etc).

\section{Data Description and Critical Dates}
\label{sec:data}

In this section, we discuss all four types of data collected and analyzed for this study (shown in Table~\ref{tab:metrics}): ($i$) traffic volume data, ($ii$) daily cases number including daily confirmed cases and daily death number, ($iii$) weather information, ($iv$) social distancing -related data, and ($v$) crash data. For each category of data, we then elaborate on the corresponding data source and the embodied attributes, respectively.

\noindent \textbf{\\Timeline of shotdown and data collection period.}
Knowing these specific dates are important for researchers to explore the changes in overall traffic volume patterns across Detroit pertaining to COVID-19 and the impact of these four types of data on the prediction of the confirmed cases. Note that in Detroit, the onset of COVID-19 was estimated to be ion {\em March 1th}, the shutdown started on {\em March 19th}, and people started going back to work in the office on {\em June 1st}. however, both, the number of closed businesses in the downtown area and the number of people working from home in general continued to remain high. 

Thus, we defined the shutdown period as going into effect from the 19th of March to 1st of June. We collected and analyzed transportation data from Detroit before the first COVID-19 confirmed cases, during the pandemic, and after the release of shutdown\,---\,from 1/1/2019 to 6/30/2020\,---\,covering more than one and a half years. To keep data consistent, we also analyzed daily cases number, weather information, and social index data that spans the same time period.

\subsection{Traffic Volume Data}
We collected and analyzed transportation data from 73 signalized intersection sites with advanced Remote Traffic Signal Management System (RTSMS) Level-II. Those locations have continuous data collection and analytics metrics through camera detection. They are all owned by the City of Detroit out of a total of 787 City-owned signals and another approximately 700 signals owned by other jurisdictions around the City. These signalized intersections are indispensable parts of urban traffic networks since around two-thirds of urban vehicle miles traveled on signal controlled roads \cite{mccracken1996demonstration}. Fig.~\ref{fig:MiovisionLocations} depicts the geographical distribution of the studied 73 signalized intersections, which are highlighted by the green dots. Those locations provided aggregated daily traffic volume data for the following 10 attributes\,---\,Bus, MotorizedVehicle, PickupTruck, ArticulatedTruck, SingleUnitTruck, Pedestrian, Motorcycle, Car, WorkVan, and Bicycle. Those attributes were later compounded into 6 as described in future sections. Additionally, it must be noted that in 2019, the number of Level-II RTSMS locations was limited to 25. However, to account for that discrepancy the normalized average volume per intersection was used.

\begin{figure}[ht]
\centering
\includegraphics[scale=0.28]{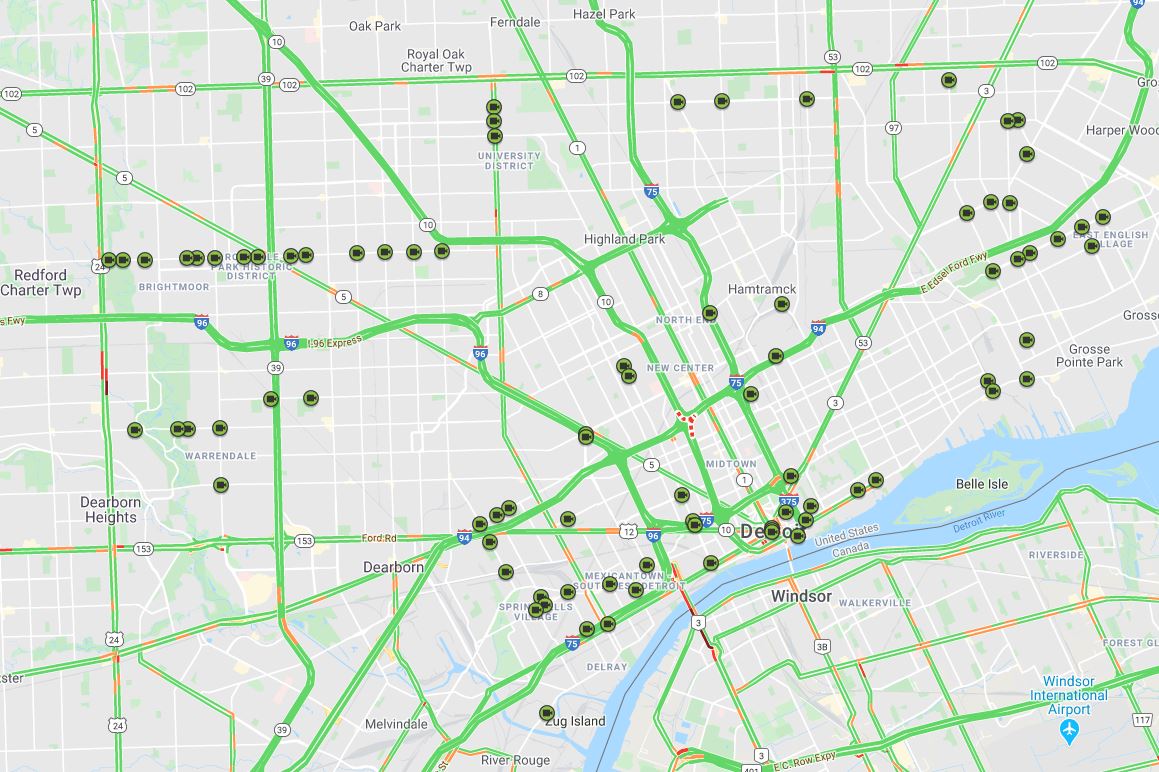}
\vspace{-0.3cm}
\caption{The distribution of 73 intersections in Detroit.}
\label{fig:MiovisionLocations}
\end{figure}

\subsection{Daily Cases}

We obtained and analyzed the daily cases data from Michigan's official Coronavirus dashboard$\footnote{https://www.michigan.gov/coronavirus/0,9753,7-406-98163\_98173---,00.html}$, including ($i$) the number of confirmed cases, and ($ii$) the number of reported deaths. More specifically, as to the number of confirmed cases, the number refers to the disease onset date; otherwise, either the specimen collection date of the first positive COVID-19 test or referral date is used. AS for the number of reported deaths, the corresponding value represents the actual reported date of death, and 8 confirmed deaths did not have a valid date available and are not included in the collected data.

\begin{figure}[ht]
\centering
\includegraphics[scale=0.62]{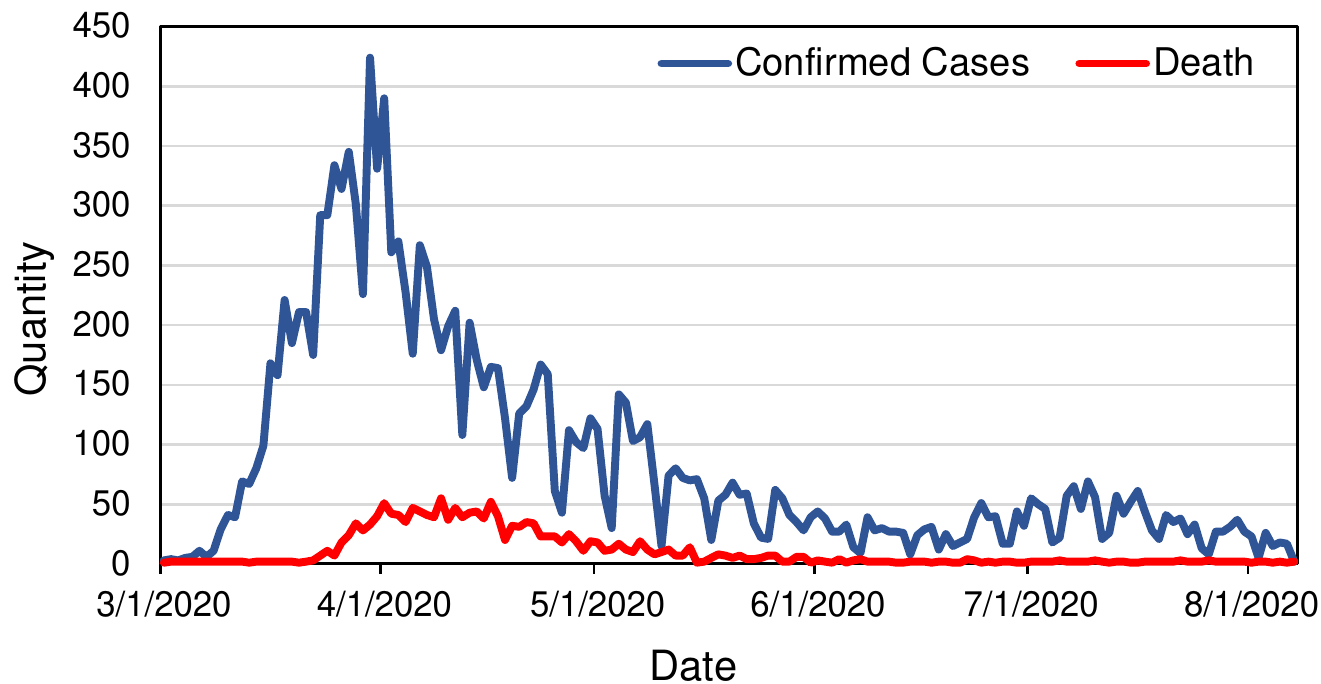}
\vspace{-0.2cm}
\caption{The number of daily confirmed cases and death in Detroit.}
\label{fig:MiovisionLocations}
\vspace{-0.4cm}
\end{figure}

\subsection{Weather Data}
Since previous work research that weather factors \eg temperature (°C) and the wind speed (mph) may be a contributor to the spread of COVID-19, we also included weather information as one of the input for COVID-19 prediction. We collected weather data from the official website of National Oceanic and Atmospheric Administration$\footnote{https://www.usa.gov/federal-agencies/national-weather-service}$, and analyzed six weather-related attributes including: Rain precipitation, Snow precipitation, Average temperature,  Maximum temperature, Minimum temperature, and average wind speed.


\subsection{Social Distancing Information}

We also collected and analyzed the social distancing related attributes for Wayne County and Michigan State from the COVID-19 Impact Analysis Platform$\footnote{https://data.covid.umd.edu/}$ published by the University of Maryland. This data was not granular enough to account for specifically the City's boundaries, however, it included valuable data for the various input contributing to social distancing on both the County and State levels.

In particular, the social distancing index is calculated from the six mobility indicators by the following equation: social distancing index = 0.8 $\times$ [\%staying home + 0.01 $\times$ (100 $-$ \%staying home) $\times$ (0.1 $\times$ \%reduction of all trips compared with pre-COVID-19 benchmark $+$ 0.2 $\times$ \%reduction of work trips $+$ 0.4 $\times$ \%reduction of non-work trips $+$ 0.3 $\times$ \%reduction of travel distance)] $+$ 0.2 $\times$ \%reduction of out-of-county trips. The choice of weight is based on the shared travel ratio of residents and tourists (\eg about 20\% of all trips are trips outside the county, which leads to a choice of 0.8 for residents and 0.2 for out-of-county travel); what trips are considered more important (\eg work travel is more important than non-work travel). A higher social distance index score refers to the fewer chances for close interpersonal interaction and reduced opportunities for the transmission of COVID-19.

\subsection{Crash Data}
To figure out the impact of COVID-19/social distancing metrics on the rate and severity of crashes, we also collected 19 crash metrics for Detroit from the Michigan State Police (MSP) Traffic Crash Reporting System (TCRS)$\footnote{https://milogintp.michigan.gov/mdot-waps6/crash/}$, which contains the information related to the number of total crashes per day broken down by severity, other reasons, and types of crashes. Table~\ref{tab:metrics} lists these metrics. The monitoring frequency for all metrics are per day.

\begin{table*}[b]
\caption{Analyzed Metrics Summary.}
\label{tab:metrics}
\vspace{-0.1cm}
\centering
\scalebox{0.81}{
\begin{tabular}{|m{2cm}|m{0.25cm}|m{6cm}|m{9cm}|}
\hline
\multicolumn{1}{|c|}{\textbf{Data Category}} & \multicolumn{1}{c|}{\textbf{\# }} & \multicolumn{1}{c|}{\textbf{Metrics}} & \multicolumn{1}{c|}{\textbf{Definition}} \\ \hline
\textbf{Traffic Volume} & \multicolumn{1}{c|}{10} & Bus, MotorizedVehicle, PickupTruck, ArticulatedTruck, SingleUnitTruck, Pedestrian, Motorcycle, Car, WorkVan, and Bicycle. & Traffic volume by classification: number of each transportation mode per day. \\ \hline
\textbf{Daily Case} & \multicolumn{1}{c|}{2} & Confirmed cases, Confirmed death & Number of confirmed cases and death per day. \\ \hline
\multicolumn{1}{|c|}{\multirow{4}{*}{\textbf{Weather Data}}} & \multicolumn{1}{c|}{\multirow{4}{*}{6}} & Rain precipitation, Snow precipitation & Volume of precipitation of rain and snow per day. \\ \cline{3-4} 
 &  & Average temperature, Maximum temperature, Minimum temperature & Value of average/maximum/minimum temperature per day. \\ \cline{3-4} 
 &  & Wind speed & Average Speed of wind per day. \\ \hline
\multicolumn{1}{|c|}{\multirow{8}{*}{\textbf{Crash}}} & \multicolumn{1}{c|}{\multirow{8}{*}{19}} & Total crashes & Number of total crashes per day. \\ \cline{3-4} 
 &  & Fatal, Serious, Minor, Possible, None & Severity of crashes (worst injury). \\ \cline{3-4} 
 &  & Ped, Cyclist, YoungDriver(Under 24) & Other reason for crash (Crashes with non-motorized roadway users and young drivers under 24 years of age). \\ \cline{3-4} 
 &  & Single motor vehicle, Head on, Head on left turn, Angle, Rear end, Rearend right turn, Sideswipe same, Sideswipe opposite, Backing, Other, UnknownNull or not entered & Type of crashes.\\ \hline
\multirow{21}{*}{\textbf{\begin{tabular}[c]{@{}c@{}}Social \\ Distancing \\ Related Data\end{tabular}}} & \multicolumn{1}{c|}{\multirow{21}{*}{21}} & Social distancing index\ & An  integer  from  0$\backsim$100. The higher value indicates a higher level social distancing. \\ \cline{3-4} 
 &  & \% staying home\ & Percentage of residents staying at home. \\ \cline{3-4} 
 &  & Trips/person\ & Average number of all trips taken per person per day. \\ \cline{3-4} 
 &  & \% out-of-county trips\ & Percentage of all trips that cross county borders. \\ \cline{3-4} 
 &  & \% out-of-state trips\ & Percentage of all trips that cross state borders. \\ \cline{3-4} 
 &  & Miles/person\ & Average person-miles traveled on all modes per person per day. \\ \cline{3-4} 
 &  & Work trips/person\ & Number of work trips per person per day. \\ \cline{3-4} 
 &  & Non-work trips/person\ & Number of non-work trips per person per day. \\ \cline{3-4} 
 &  & New COVID cases\ & Number of COVID-19 daily new cases. \\ \cline{3-4} 
 &  & \% change in consumption\ & \% change in consumption from the pre-pandemic baseline. \\ \cline{3-4} 
 &  & COVID exposure/1000 people\ & Number of residents already exposed to coronavirus per 1000 people. \\ \cline{3-4} 
 &  & Unemployment claims/1000 people\ & New weekly unemployment insurance claims/1000 workers. \\ \cline{3-4} 
 &  & Unemployment rate\ & Unemployment rate updated weekly. \\ \cline{3-4} 
 &  & \% working from home\ & Percentage of workforce working from home. \\ \cline{3-4} 
 &  & COVID death rate\ & \% deaths among all COVID-19 cases. \\ \cline{3-4} 
 &  & New cases/1000 people\ & Number of COVID-19 daily new cases per 1000 people. \\ \cline{3-4} 
 &  & Active cases/1000 people\ & Number of active COVID-19 cases per 1000 people. \\ \cline{3-4} 
 &  & \#days: decreasing COVID cases\ & Number of days with decreasing COVID-19 cases. \\ \cline{3-4} 
 &  & Testing capacity\ & Ability to provide enough tests. \\ \cline{3-4} 
 &  & Tests done/1000 people\ & Number of COVID-19 tests already completed per 1000 people. \\ \cline{3-4} 
 &  & \% ICU utilization\,Imported COVID cases\ & \% ICU unites occupied with COVID-19 patients. \\ \hline
\end{tabular}}
\end{table*}

\section{Statistical Analysis}
\label{sec:statistic}
In this section, we conduct statistical analysis on the collected metrics and we present our interesting observations related to the changes of traffic volume and pattern, correlation studies, and crashes statistics analysis.

\subsection{Changes of Traffic Volume and Pattern}

\subsubsection{\bf{Changes of Traffic Volume in 2020}}
\begin{figure}[h]
\centering
\includegraphics[scale=0.58]{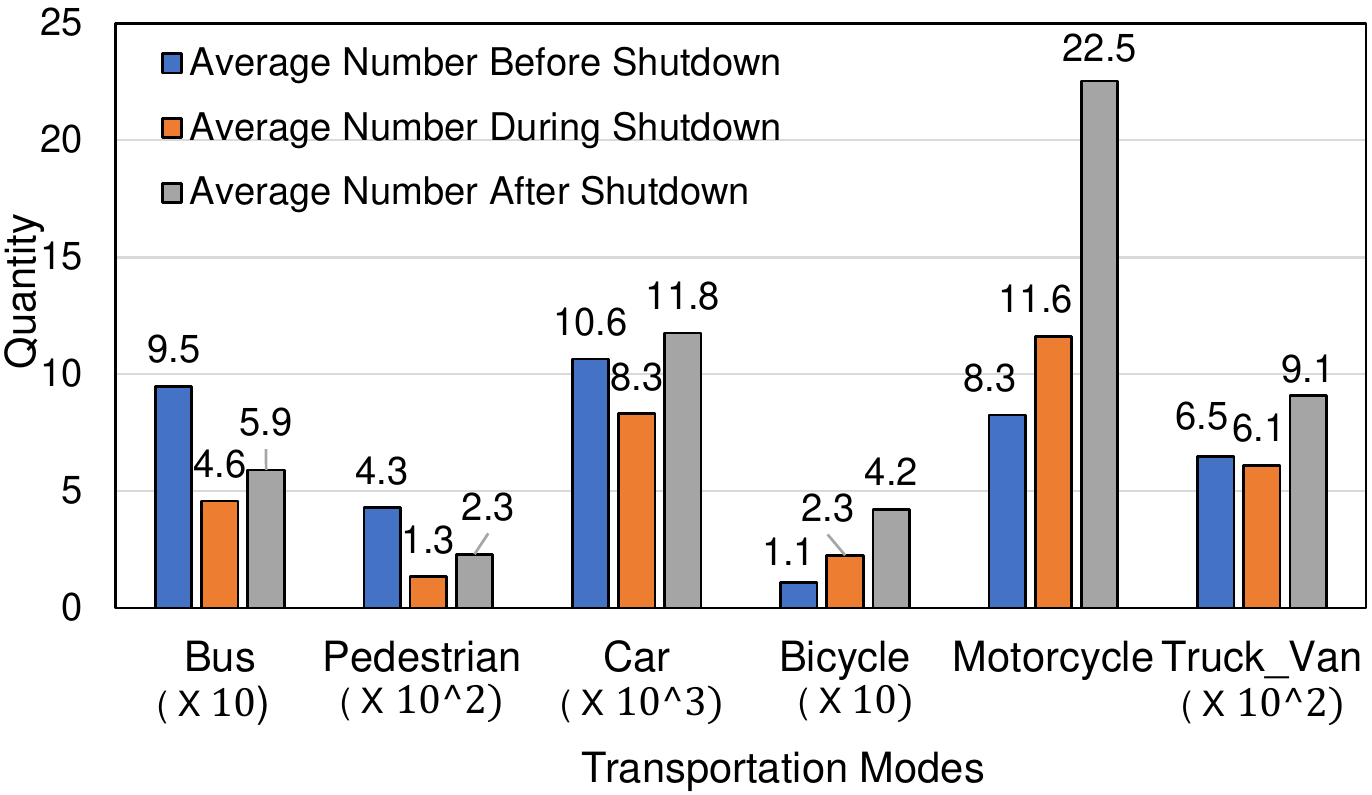}
\caption{Changes in the daily number of each transportation model before, during, and after quarantine in Detroit.}
\label{fig:traffic}
\end{figure}

We first sought to answer the question of "What's the sudden spatial traffic patterns pertaining to COVID-19". To answer this question, we first calculate the average number of each transportation mode among 73 signalized intersections for each monitoring day, then We further draw the average number of each mode during the period before, during, and after quarantine in Detroit (shown in Fig.~\ref{fig:traffic}).
Out of the 10 collected metrics related to traffic counts at signalized intersections (shown in Table~\ref{tab:metrics}), we combined some of those metrics an consolidated the classifications down to six categories. For example, the value of "Truck\_Van" represents the total number of all recorded types of trucks and vans, while the value of "Car" indicates the total quantity of motorized vehicles and traditional cars. 

It can be seen that the number of buses, pedestrians, and cars both showed declining trends during the shutdown period and then increased after the ending of the shutdown. On the contrary, it is notable that the quantity of Bicycle and Motorcycle increased sharply even after the statewide quarantine order. For example, the average number of Bicycle per day increased to approximately 2$\times$ its original value and even increase by approximately 4$\times$ during and after the quarantine, respectively. Similarly, the average number of Motorcycle surged to approximately 3$\times$ its before shutdown numbers.

\begin{tcolorbox}[left = 0mm, right = 0mm, top = 0mm, bottom = 0mm, boxsep = 1mm, arc = 0mm]
 \textbf{Observations 1:} Although cars are still the predominant transportation mode, biking and motorcycling have demonstrated a transition in usage and are showing an increased popularity (up to 4$\times$ of the previous volume) in the post-COVID-19 urban mobility for Detroit. 
 These numbers do not account for the seasonal weather impact on Bike and Motorcycle usage. 
\label{observations1}
\end{tcolorbox}

This observation may be attributed to the fact that cycling can be an alternative mode of transport as it can be compatible with social distancing regulations and allow for short individual trips. It must be notes that this excercise does not account for the seasonal weather impact on Bike and Motorcycle usage which coincided with the onset of COVID-19.  Although literature exploring the role of cycling in previous epidemics is rare, it is recognized that one of the factors leading to the rise of e-bikes in China was the 2002–2004 SARS outbreak as people tried to avoid overcrowded public transportation services \cite{weinert2007transition, simha2016disruptive}. Additionally, the same pattern was also observed in New York City \cite{teixeira2020link}, showing some evidence of a modal transfer from some subway users to the bike-sharing system.

\begin{tcolorbox}[left = 0mm, right = 0mm, top = 0mm, bottom = 0mm, boxsep = 1mm, arc = 0mm]
 \textbf{Observations 2:} The total number of trucks and vans was almost the same before and during the shutdown. An increased of approximately 40\% is observed in the post-shutdown numbers.
\label{observations1}
\end{tcolorbox}


The above truck-related observation might be attributed to: $i)$ with the outbreak of the epidemic, a portion of Detroit citizens are turning to online shopping methods. However, due to the shutdown policies, some industry-related trucks and vans stopped running, leading to a relatively stable volume of trucks and vans during the quarantine period. $ii)$ after the shutdown period, although there are no specific rules to limit the transportation of trucks and vans, it is expected that the increased delivery demand was maintained in addition to the increased demand from the reopening of industrial activity and road construction. \\


\subsubsection{\bf{Changes of Traffic Pattern between 2019 and 2020}}
\begin{figure}[h]
\centering
\includegraphics[scale=0.4]{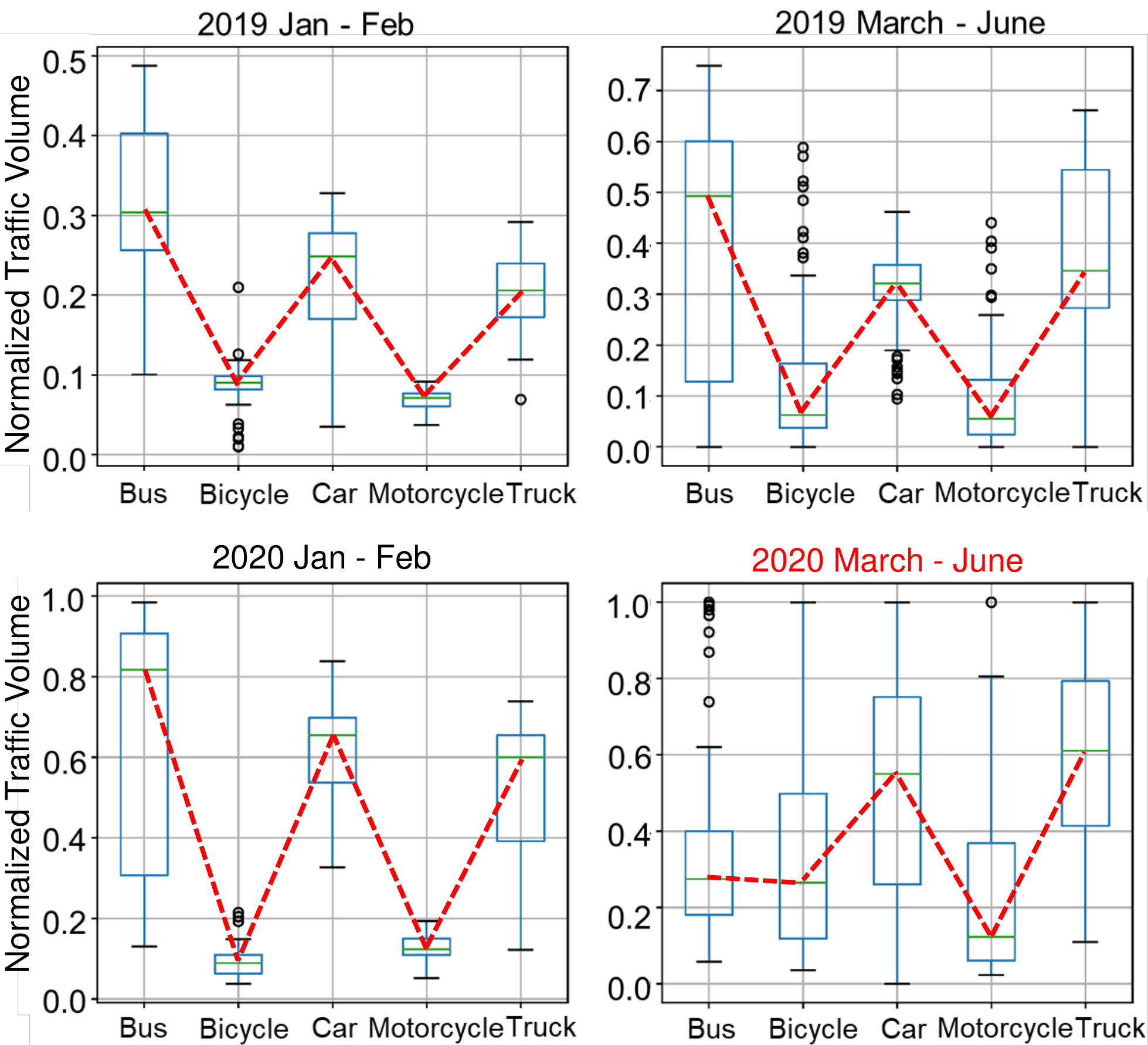}
\caption{The change of traffic patterns from 2019 to 2020 pertaining to COVID-19.}
\label{fig:TemporaryChange}
\end{figure}

In order to the answer \textit{ "how did traffic patterns change from 2019 to 2020 considering the impact of COVID-19"}, we explored the temporary distribution of each transportation mode's quantity among 2019 and 2020, including bus, bicycle, car, motorcycle, and truck. More specifically, we divided the collected traffic data set into four groups based on the temporary periods, \ie 2019 Jan through Feb, 2019 March through June, 2020 Jan through Feb, and 2020 March through June (the outbreak time period of COVID-19). For each group of data sets, we calculated the normalized traffic volume of each transportation mode and draw the corresponding box plot (Fig.~\ref{fig:TemporaryChange}) that displays the minimum, first quartile, median, third quartile, and maximum value of the normalized traffic quantities. Then, we used red straight lines to connect the medium value of each type of transportation volume, so that the shape of the red lines on a single sub-figure of Fig.~\ref{fig:TemporaryChange} provides the general traffic pattern information. In can been seen that the red lines of sub-figures titled as "2019 Jan - Feb", "2019 March - June", and "2020 Jan - Feb" both show the "$W$" shape, \ie the medium value of bicycle and motorcycle is significantly lower than the value of bus. However, when it comes to "2020 March - June", this "$W$" shape is disappeared since the medium value of bicycles is increased while the medium value of buses is declined markedly compared with other sub-figures.



\begin{tcolorbox}[left = 0mm, right = 0mm, top = 0mm, bottom = 0mm, boxsep = 1mm, arc = 0mm]
 \textbf{Observations 3:} Comparing the traffic volume data of 2019 and 2020, the traffic pattern of 2020 March - June (outbreak period of COVID-19) is significantly different from the patterns of 2019 and the first two months of 2020.
\label{observations1}
\end{tcolorbox}

\subsection{Correlation Study}
To further explore the correlation between each type of collected metrics, such as the correlation between the number of new confirmed cases with the daily traffic volume, we conduct correlation coefficient analysis and joint distribution analysis, which present a statistic analysis results on their correlation.

\subsubsection{\bf{Correlation Coefficients}}
One of the more frequently reported statistical methods involves correlation analysis where a correlation coefficient is reported representing the degree of linear association between two variables \cite{taylor1990interpretation, lee1988thirteen}.

In this work, we calculated the correlation coefficients between $i)$ transportation modes (\eg bus, pedestrian, bicycle, car, motorcycle, and truck), $ii)$ total crashes, $iii)$ weather info (\eg average temperature, rain precipitation,  and daily average wind speed), $iv)$ social distancing index of Wayne County and Michigan, and $v)$ daily cases (\eg daily confirmed cases and daily deaths). The formula for the correlation coefficient is defined as follows.

\begin{equation}
    \label{Eq:correlation}
    \begin{aligned}
    \rho_{ x,y }= \frac{E\left [ XY \right ]-E\left [ X \right ]E\left [ Y \right ]}{\sqrt{E\left [ X^2 \right ]-\left (  E\left [ X \right ] \right )^2 }\sqrt{E\left [ Y^2 \right ]-\left (  E\left [ Y \right ] \right )^2}  }
    \end{aligned}
\end{equation}

The correlation coefficient is a statistic that measures the linear correlation between two attributes $X$ and $Y$, with the value range of [$-1$, $+1$]. A value of $+1$ is the total positive linear correlation, 0 is no linear correlation, and $-1$ is the total negative linear correlation. The higher correlation coefficient represents the higher correlation between the two attributes.

\begin{figure}[h]
\centering
\includegraphics[scale=0.42]{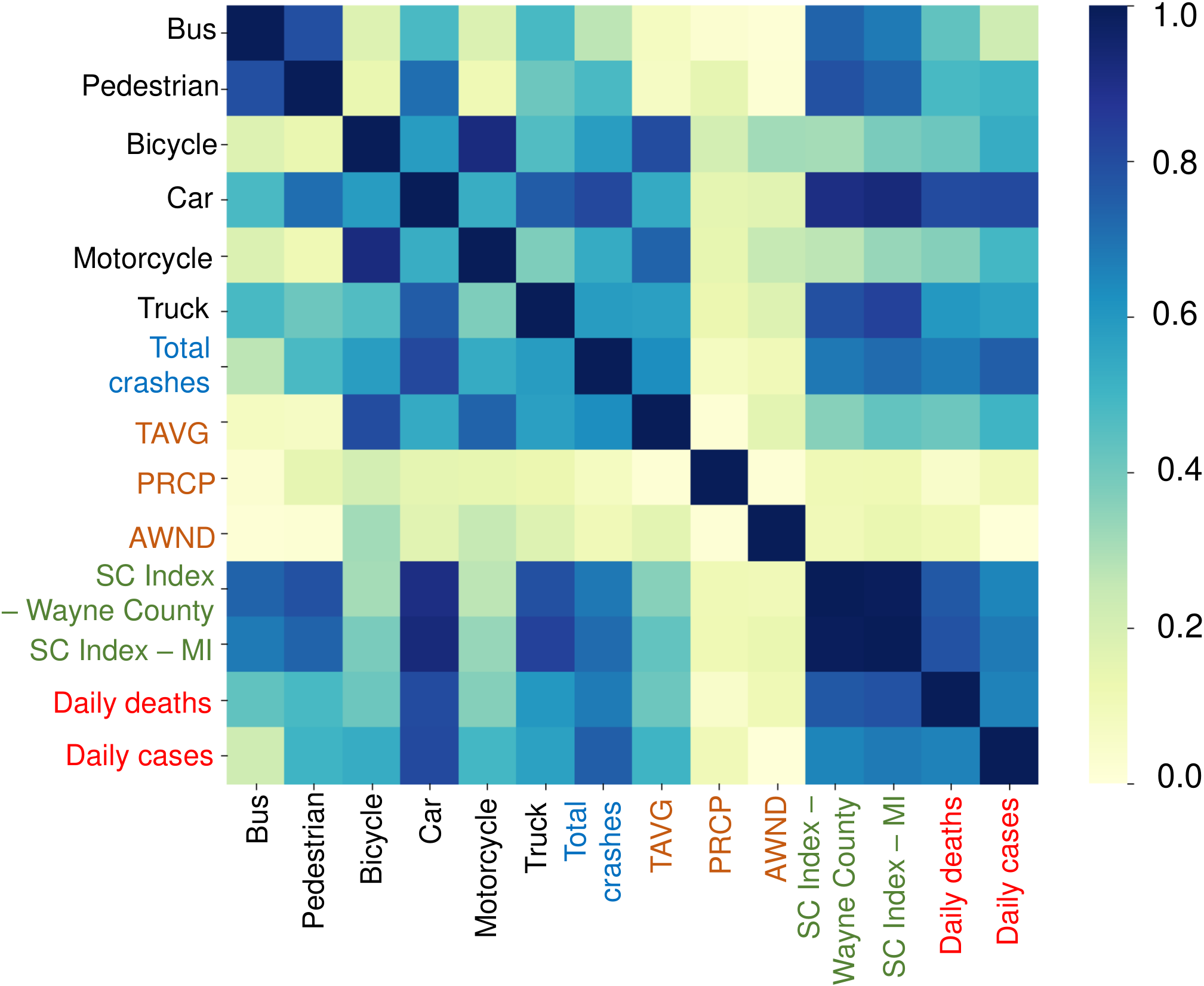}
\vspace{-0.3cm}
\caption{The absolute value of correlation coefficients. TAVG represents the average temperature per day. PRCP indicates the daily rain precipitation. AWND stands for the daily average wind speed, and SC index represents the social distancing index.}
\label{fig:CorrelationCoefficient}
\end{figure}

Fig.~\ref{fig:CorrelationCoefficient} presents the absolute values of the calculated correlation coefficients based on Equation~\ref{Eq:correlation}.  We use the gradient color from yellow to blue to indicate the lower to the higher correlation between two attributes.

\begin{tcolorbox}[left = 0mm, right = 0mm, top = 0mm, bottom = 0mm, boxsep = 1mm, arc = 0mm]
 \textbf{Observations 4: Daily cases \ie daily confirmed cases and daily death is highly related, with: \\
 $1-$ the number of transportation volume, especially for the cars, \\
 $2-$ total crashes, \\
 $3-$ social distancing index at the Wayne County level and Michigan level, and \\
 $4-$ the average temperature}.
\label{observations1}
\end{tcolorbox}

Note that although previous work revealed the wind speed (mph) is one of the factors that triggered the spread of COVID-19 \cite{csahin2020impact}, in this work, we do \textit{not}
find the high \textit{linear} correlation between the number of daily cases with the wind speed and other weather factors such as rain precipitation in Detroit based on our collected dataset.\\

\subsubsection{\bf{Joint Distribution}}
Based on the correlation coefficients, we can roughly know the correlation degree between each pair of attributes, but we cannot decisively derive the specific reason for that relationship. Therefore, we calculated and drew the joint distributions \cite{sklar1973random} for select pairs of attributes. This demonstrates the intuitive quantitative relationship between variables (linear / non-linear, or whether there is a more obvious correlation). Most importantly, the joint distributions allow us to identify the relationship between multiple attributes.

To be certain, for $\left \{ X=x,Y=y \right \} $, we found all elements in the sample space that satisfy these two values. These elements formed a subset of the sample space, and the probability of this subset was the joint probability of $\left [ P\left ( X=x,Y=y \right )  \right ] $. $\left [ p\left ( x,y \right ) =P\left ( X=x,Y=y \right )  \right ] $ is called joint PMF (joint probability mass function). The joint probability can be regarded as the probability when two events occur at the same time. Event A is $\left [ X=x \right ] $, and event B is $\left [ Y=y \right ] $, which is $\left [ P\left ( A\bigcap B \right )  \right ] $.


\begin{figure}[h]
\centering
\includegraphics[scale=0.62]{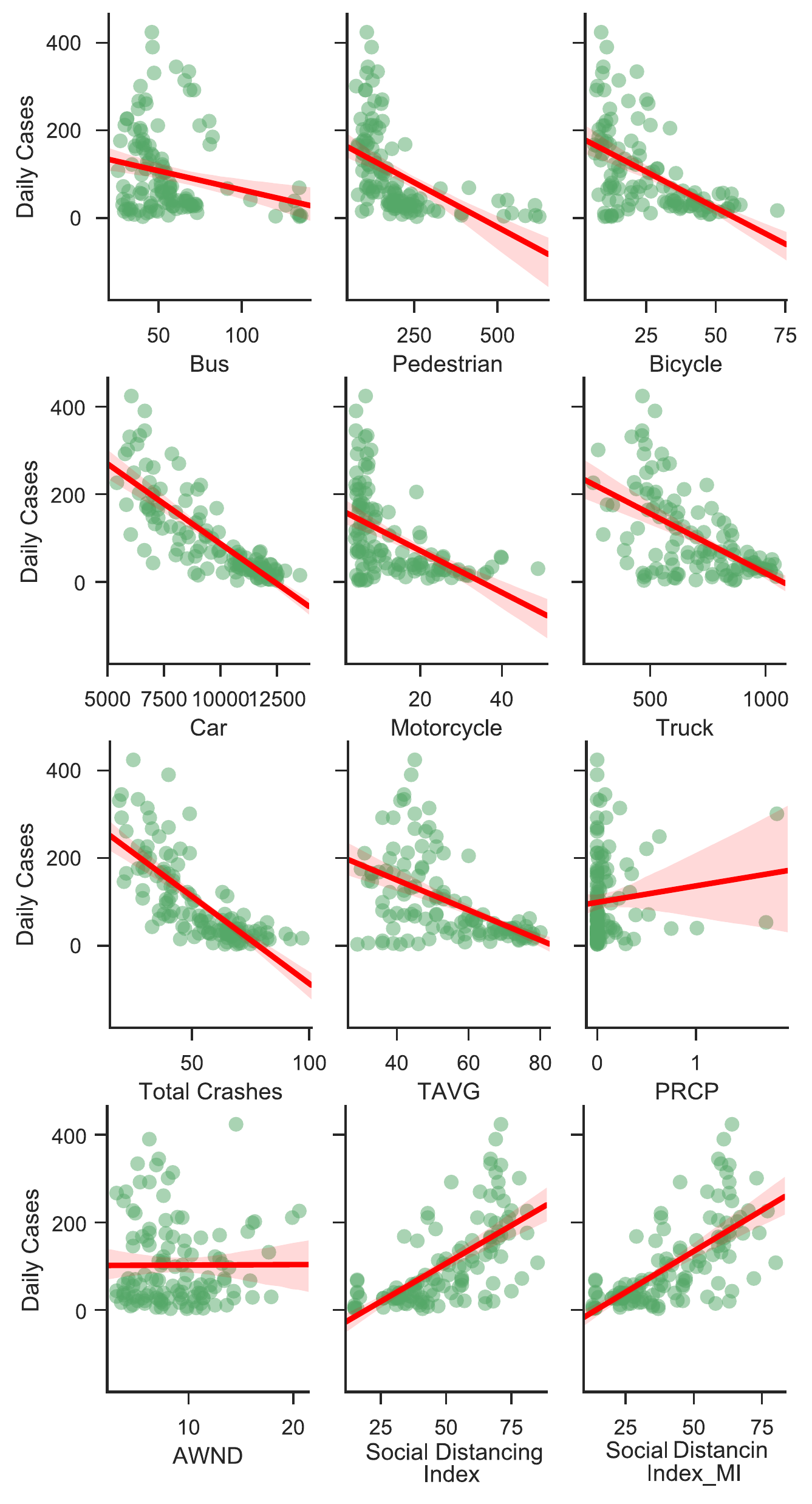}
\vspace{-0.3cm}
\caption{Joint distribution of various relevant attributes.}
\label{fig:Jointdistribution}
\vspace{-0.3cm}
\end{figure}

Fig.~\ref{fig:Jointdistribution} presents the joint distribution between the daily confirmed cases with other attributes such as transportation volume, total crashes, weather, and social distancing index from 2020 March to June. Green points display the specific value of the selected attributes, and red straight lines are used to represent the linear fit results between the pairs of two attributes, \eg the linear fit result of the daily confirmed case and the number of buses per day. The greater the slope of the red line, the stronger the linear relationship between the two attributes.

Based on the joint distribution of the daily confirmed cases and the rain precipitation shown in Fig.~\ref{fig:Jointdistribution}, we can deduce the reason why they are \textit{not linearly correlated}: no matter how large the value of daily cases is, the value of PRCP is very small, \ie the rainfall in Detroit from March to June of 2020 is very small, resulting in a very small range of rainfall. Under these circumstances, the value of the correlation coefficient between the daily cases and PRCP is low, indicating a low \textit{linear correlation}.

\subsection{Crashes Statistics}
\begin{figure}[h]
\centering
\includegraphics[scale=0.32]{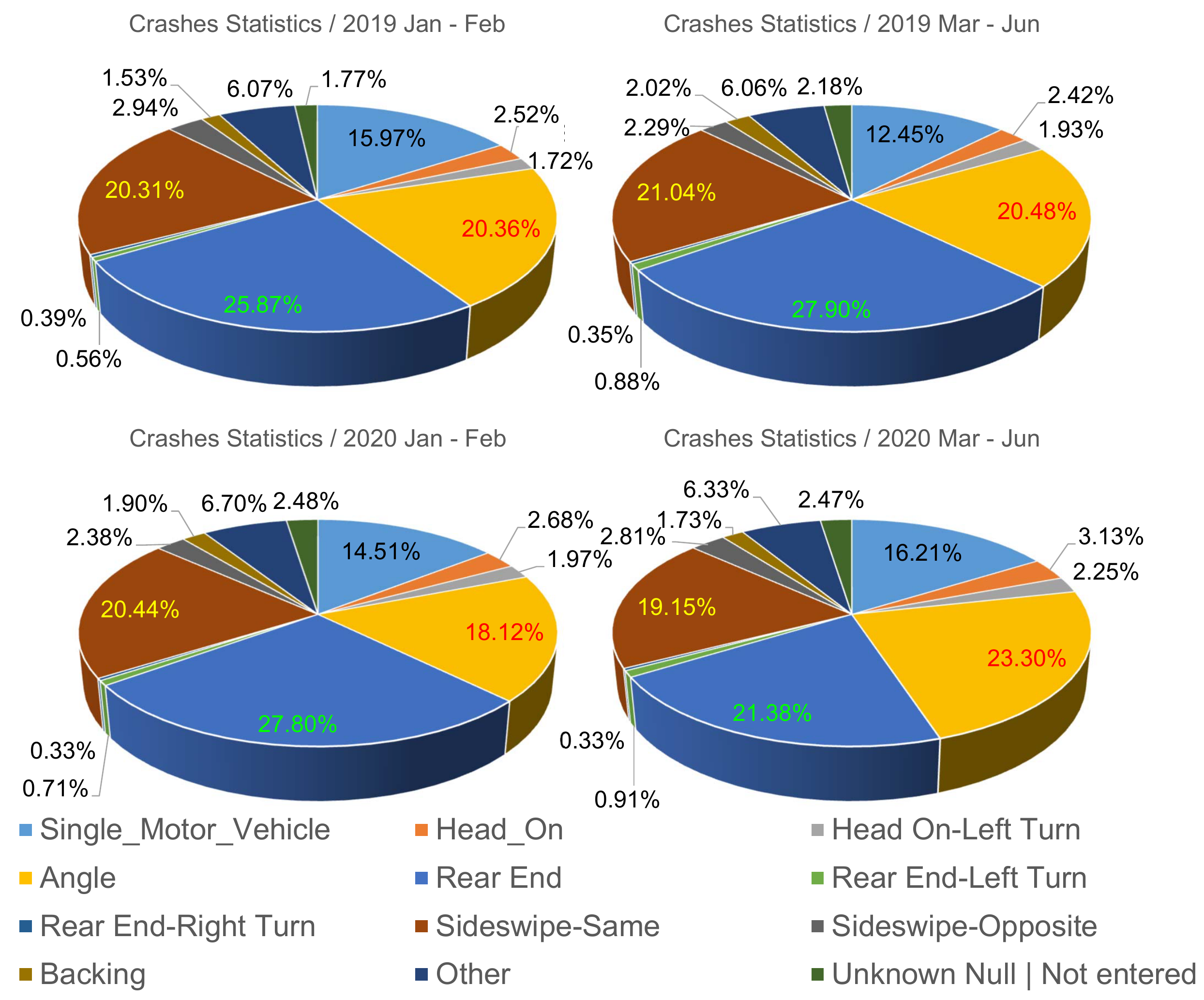}
\vspace{-0.3cm}
\caption{Distribution by crash type percentage for 2019 to 2020.}
\label{fig:CrashesStatistics}
\end{figure}

We then analyzed crash data covering 2019 Jan to Feb, 2019 Mar to Jun, 2020 Jan to Feb, and 2020 Mar to Jun. Our goal was to identify the percent distribution of the different crash types, which is shown in Fig.~\ref{fig:CrashesStatistics}. 

\begin{tcolorbox}[left = 0mm, right = 0mm, top = 0mm, bottom = 0mm, boxsep = 1mm, arc = 0mm]
 \textbf{Observations 5: When Comparing the 2020 crash type percentages from before the outbreak of COVID-19 with those during the pandemic period, a clear crash type distribution shift is observed: \\$1-$ Angle crashes became the most common moving up from third place (18.12\%$\sim$23.30\%), \\\&$2-$ Rearend crashes moved from first to second place (27.80\%$\sim$21.38\%), \\$3-$ Sideswipe crashes slightly decrease but maintained third place (20.44\%$\sim$19.15\%), and \\$4-$ Single vehicle crashes increased slightly and maintained forth place (14.51\%$\sim$16.21\%).}
 
 Data key: (2020 Jan-Feb $\sim$ 2020 March-June)
\label{observations1}
\end{tcolorbox}

By observing the crash type percentage distribution change, it can be seen that rear-end crashes which used to the be the most predominant crash type was overtaken by angle crashes. This observation is indicative of a potentially problematic change in driver behavior: more distraction and speeding, and less adherence to traffic rules and controls. Additionally, rear-end crashes are usually more prevalent in high-volume areas, which is not the case during COVID-19 when the volumes are much lower. A deeper dive into the changes in the total crash numbers and crash rate is needed due to the unexpected crash behavior during COVID. Those analyses were not completed or discussed as part of this paper.

\section{COVID-19 Confirmed Cases Prediction}
\label{sec:method}
In this section, we aim to conduct a further study on the influence of traffic volume data,  weather information data, crash data, and social distancing index information on the confirmed-cases prediction.Our ultimate goal is building a suitable deep learning model to predict the
number of confirmed cases based on $(i)$ traffic volume data, $(ii)$ daily cases number including daily confirmed cases and daily death number, $(iii)$ weather information, $(iv)$ social distancing-related data, and $(v)$ crash data \\
\vspace{-0.3cm}
\subsection{Problem Formulation and Solution}

\noindent \textbf{Problem Definition.} 
We formulate the problem of predicting the number of COVID-19 confirmed cases as a regression problem. Specifically, we use $T=\left \{  \text{input}_{i} \right \}_{i=1}^{n}$ to represent our training data set, in which $\text{input}_{i} \in I$ denotes all input features, \ie the 58 features present in Table~\ref{tab:metrics} of Sec.~\ref{sec:data}. Our goal is to employ the best method to learn the function $f$, which minimizes the loss function $\ell\left ( h\left ( \text{input} \right ); ground truth\right )$, a measurement of the difference between the desired output and the actual output of the current model, such that the trained model is able to predict the number of confirmed cases over a specific prediction horizon with high performance. Besides, we choose 21 days as our monitoring window, and we aim to predict confirmed-cases for the next 7 days.\\

\begin{figure}[H]
\centering
\includegraphics[scale=0.55]{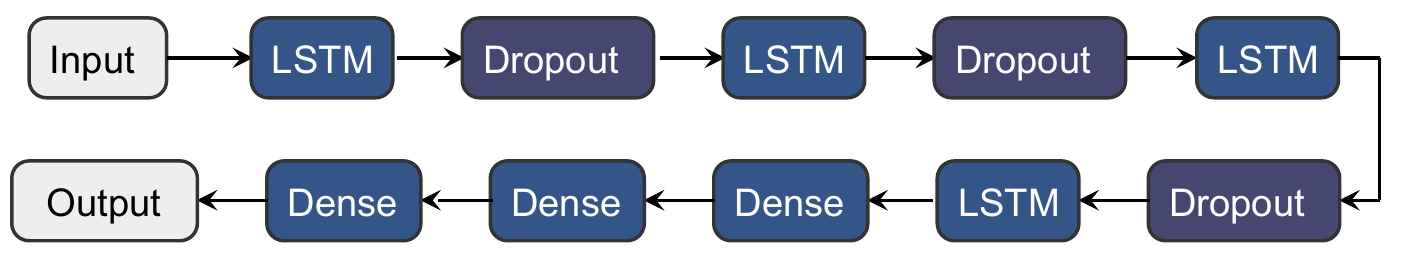}
\vspace{-0.3cm}
\caption{The structure of the LSTM.}
\label{fig:lstm}
\end{figure}

\noindent \textbf{Deep Learning Model Selection.} Recently, machine learning methods have been applied with success in regression tasks. We tackle the confirmed-cases prediction problem using Long Short-Term Memory Networks (LSTMs) \cite{hochreiter1997long,dos2017predicting} since it has become highly successful learning models for both classification and regression problems across diverse domains \cite{basak2019analyzing, hong2019fault, lu2020making, lu2019collaborative}. Specifically, LSTM is a type of recurrent neural networks (RNNs) with the capability of processing sequences of sequential data sets. After being proposed by Hochreiter and Schmidhuber \cite{hochreiter1997long}, LSTM has been proved the ability to address long-term back-propagating issues. It includes a memory cell and a gating mechanism, which allows it to decide what is kept in the memory cell, and how the new input data contributes to what is already in the memory cell. Fig.~\ref{fig:lstm} depicts the structure of the LSTM model that we deployed for the confirmed-cases prediction.

\noindent \textbf{Effective Measurements.}

To be able to design the best prediction method, we need some metrics to accurately measure the wellness of our prediction approaches. To begin with, we use some commonly-used measures for our study:
coefficient of determination ($R^2$), mean square error (MSE), and the root mean square error (RMSE), which both are the commonly used evaluation metrics for the regression problem \cite{anastassopoulou2020data}. $R^2$ score is widely used to indicate the fit of the machine learning model, \ie the higher value, the better fit result generated by the model. The maximum value of $R^2$ is 1 (ideal case), and it may be a negative value with a range of $(-\infty, 1]$.  MAE measures the average magnitude of the errors in a set of predictions, without considering their direction. It’s the average over the test sample of the absolute differences between prediction and actual observation where all individual differences have equal weight. RMSE is a quadratic scoring rule that also measures the average magnitude of the error. It’s the square root of the average of squared differences between prediction and actual observation.

Suppose the input data (ground truth) is $y=\left \{ y_{1},y_{2},...,y_{N} \right \}$, and the prediction result is noted as $\quad \hat{y}=\left \{  \hat{y}_{1},\hat{y}_{2},...,\hat{y}_{N}\right \}$. MSE and RMSE are defined as: 

\begin{equation}
    \label{Eq:mae}
    \begin{aligned}
    MAE=\frac{1}{N} \sum_{i=1}^{N} \left | \hat{y}_{i}-y_{i}  \right | 
    \end{aligned}
\end{equation}

\begin{equation}
    \label{Eq:rmse}
    \begin{aligned}
    RMSE =\sqrt{MSE}=\sqrt{\mathrm{\frac{1}{N}\sum_{i=1}^{N}  \left ( y_{i}-\hat{y}_{i}   \right ) ^{2} } }
    \end{aligned}
\end{equation}

Specifically, the formula to calculate $R^2$ is defined as follows:

\begin{equation}
    \label{Eq:R2}
    \begin{aligned}
    R^2=1-\frac{RSS}{TSS} =1-\frac{\sum_{i=1}^{N}\left ( y_{i}-\hat{y}_{i}  \right )^2  }{\sum_{i=1}^{N}\left ( y_{i}-\bar{y}   \right )^2  } 
    \end{aligned}
\end{equation}

\begin{equation}
    \label{Eq:y_mean}
    \begin{aligned}
    \bar{y}=\frac{1}{N} \sum_{i=1}^{N}  y_{i}
    \end{aligned}
\end{equation}

Where TSS (total sum of squares) is the difference between all samples and the mean value, which is $N \times$ of the variance. Besides, RSS (residual sum of squares) is the sum of the squares of all sample errors, which is $N \times$ times the MSE. When the predicted value of all samples is the same as the true value, RSS is 0, so $R^2$ equal to 1 (ideal case).

\subsection{Model Creation}
In this subsection, we introduce the experimental hardware and the used packages, then we give a detailed explanation of why we conduct experiments on six experimental groups, and what's the precise format of our experimental input.\\

\noindent \textbf{Experimental Setup.}
In this work, we adopt NVIDIA GPU Workstation as our experiment platform, which is powerful hardware with the high-quality components (4$\times$GeForce RTX 2080 Ti graphics cards) with Intel Xeon E5-2690 v4 (CPU), 2.6 GHz of frequency, 14 cores, 64 GB memory, and installed with Ubuntu 16.04.6 LTS (operating system). NVIDIA GPU Workstation is capable of delivering the cluster-level performance for even the demanding applications \cite{spiga2012phigemm, morozov2011molecular}. The models learned in this paper are implemented in Python, using TensorFlow 1.13.1 \cite{abadi2016tensorflow}, Keras 2.1.5 \cite{gulli2017deep}, and scikit-learn libraries \cite{pedregosa2011scikit} for model building. \\


\noindent \textbf{Experimental Groups.} To show the impact of traffic volume data, weather-related metrics, crash data, and social distancing related data on the confirmed-cases prediction, we conduct experiments on six experimental groups. Our first step is to combine all five categories of 58 features present in Table~\ref{tab:metrics} of Sec.~\ref{sec:data} to train models using LSTM methods, and we label this group as A Group (A represent all). Then, we exclude all traffic volume metrics but keep the left features, and we denote it as A-T Group. Similarly, we exclude weather information but keep other features, and we get the A-W Group. Since we have two levels of social distancing index, \ie social distancing index of Wayne County (denoted as SD), and social distancing index of Michigan state (marked as SDM), we delete SD and SDM to get A-SD group and A-SDM group, respectively. Finally, in order to figure out the impact of crashes data (noted as C) on the confirmed-cases prediction, we get A-C group. Table~\ref{tab:input} shows the input features for A Group, A-T Group, A-W Group, A-SD Group, A-SDM, and A-C Group.

\begin{table}[h]
\vspace{-0.5cm}
\caption{Input features for six experimental groups.}
\label{tab:input}
\centering
\scalebox{0.6}{
\begin{tabular}{|c|c|c|c|c|c|c|c|}
\hline
\textbf{Group} & \begin{tabular}[c]{@{}c@{}}\# of\\ Metrics\end{tabular} & \begin{tabular}[c]{@{}c@{}}Traffic\\ Volume \end{tabular} & \begin{tabular}[c]{@{}c@{}}Daily\\ Case\end{tabular} & \begin{tabular}[c]{@{}c@{}}Weather\\ Data\end{tabular} & Crash & \begin{tabular}[c]{@{}c@{}}Social\\ Distancing\\ Related\\ Data\end{tabular} & \begin{tabular}[c]{@{}c@{}}Social \\ Distancing\\ Related\\ Data\_MI\end{tabular} \\ \hline
\textbf{A} & 79 & \textbf{$\surd$} & \textbf{$\surd$} & \textbf{$\surd$} & \textbf{$\surd$} & \textbf{$\surd$} & \textbf{$\surd$} \\ \hline
\textbf{A-T} & 69 & {\color[HTML]{FE0000}\textbf{×}} & \textbf{$\surd$} & \textbf{$\surd$} & \textbf{$\surd$} & \textbf{$\surd$} & \textbf{$\surd$} \\ \hline
\textbf{A-W} & 73 & \textbf{$\surd$} & \textbf{$\surd$} & {\color[HTML]{FE0000}\textbf{×}} & \textbf{$\surd$} & \textbf{$\surd$} & \textbf{$\surd$} \\ \hline
\textbf{A-SD} & 58 & \textbf{$\surd$} & \textbf{$\surd$} & \textbf{$\surd$} & \textbf{$\surd$} & {\color[HTML]{FE0000}\textbf{×}} & \textbf{$\surd$} \\ \hline
\textbf{A-SDM} & 58 & \textbf{$\surd$} & \textbf{$\surd$} & \textbf{$\surd$} & \textbf{$\surd$} & \textbf{$\surd$} & {\color[HTML]{FE0000}\textbf{×}} \\ \hline
\textbf{A-C} & 60 & \textbf{$\surd$} & \textbf{$\surd$} & \textbf{$\surd$} & {\color[HTML]{FE0000}\textbf{×}} & \textbf{$\surd$} & \textbf{$\surd$} \\ \hline
\end{tabular}}
\vspace{-0.4cm}
\end{table}

\subsection{Training and Validation Methodology}

\noindent \textbf{\\Training and Validation Methodology.}
Next, we provide a high-level description of our methodology before delving into the details. We use 5-fold cross-validation\cite{kohavi1995study}, which is a validation technique to assess the predictive performance of machine learning models, judge how models perform to an unseen data set (testing data set)~\cite{rodriguez2010sensitivity} and avoid the over-fitting issue during the training phase. More specifically, our data set is randomly partitioned into five equal-sized sub-samples. At a time, we take one sub-sample as the testing data set, and take the remaining four sub-samples as the training data set. Then, we fit a model on the training data set, evaluate it on the testing data set, and calculate the evaluation scores. After that, we retain the evaluation scores and discard the current model. The process is then repeated five times with different combinations of sub-samples, and we use the average of the evaluation scores as the final result for each method. 

First, we need to determine the hyperparameters of our models---an important aspect of building effective deep learning models. To be concrete, we use hold-out method \cite{kim2009estimating} to split up our training phase data set further into the parameter training process and the validation process (80\% and 20\% of the training phase data set respectively), and the validation is an unbiased evaluation of a model fit on the training dataset when tuning parameters. Then, we conduct a grid search on these values of parameters to find the best combination that achieves the highest performance. 

\begin{figure}[ht]
\centering
\includegraphics[scale=0.76]{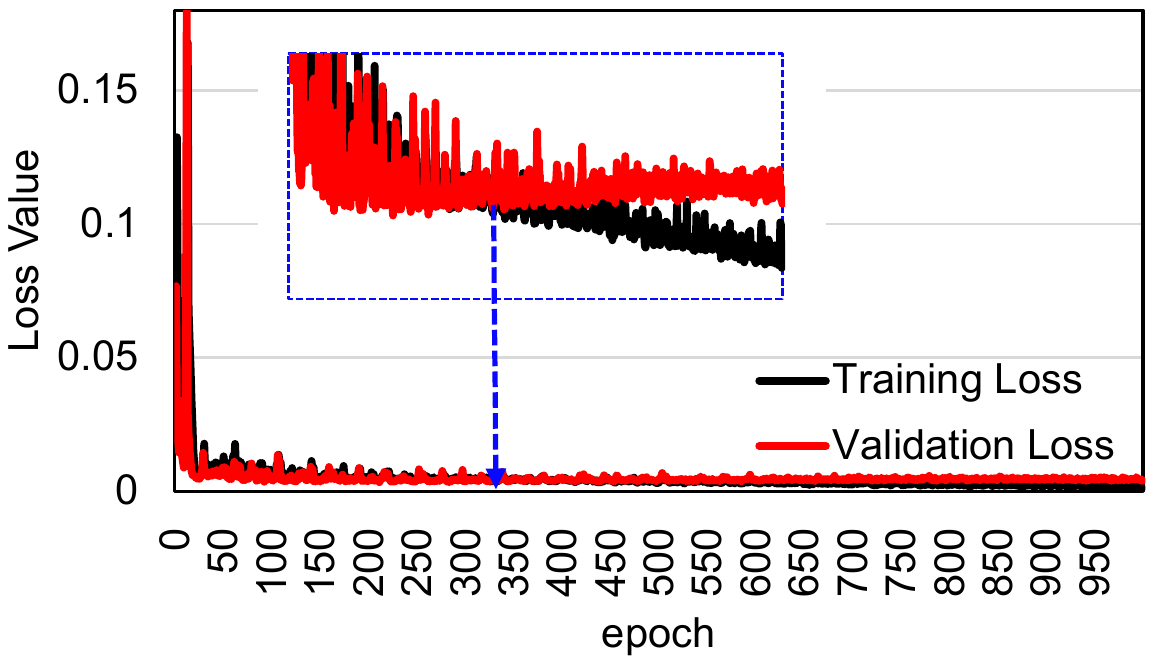}
\vspace{-0.5cm}
\caption{The validation loss reaches minimum value at 340.}
\vspace{-0.2cm}
\label{fig:epoch}
\end{figure}

\noindent \textbf{\\Avoiding Overfitting of the Models.}
Another important factor during the training process is epoch \cite{graves2005framewise}, which indicates the number of iterations of processing the input data set during the training process. With a higher value of epoch, the error on training data will reduce further; however, at a crucial tipping point, the network begins to over-fit the training data. Hence, finding the best value of the epoch is essential to avoid overfitting. Figure~\ref{fig:epoch} shows the change in the value of the training and validation loss functions (the smaller, the better) as the epoch increases. Initially, the values of the two loss functions are decreasing with higher epoch; but after 340 epochs, the value of validation loss function slowly increases (higher than the training loss), which indicates the over-fitting issue. Therefore, we choose 340 epochs for LSTM.

\section{Results and Discussion}
\label{sec:results}

In this section, we present and analyze the sensitivity of LSTM toward different feature groups. Our discussion includes supporting evidence and reasoning to explain observed trends, and implications of observed trends for the authorities and decision-makers on taking specific measures for Detroit. 

\subsection{Prediction Results and Ground Truths}
\begin{figure}[h]
\centering
\includegraphics[scale=0.34]{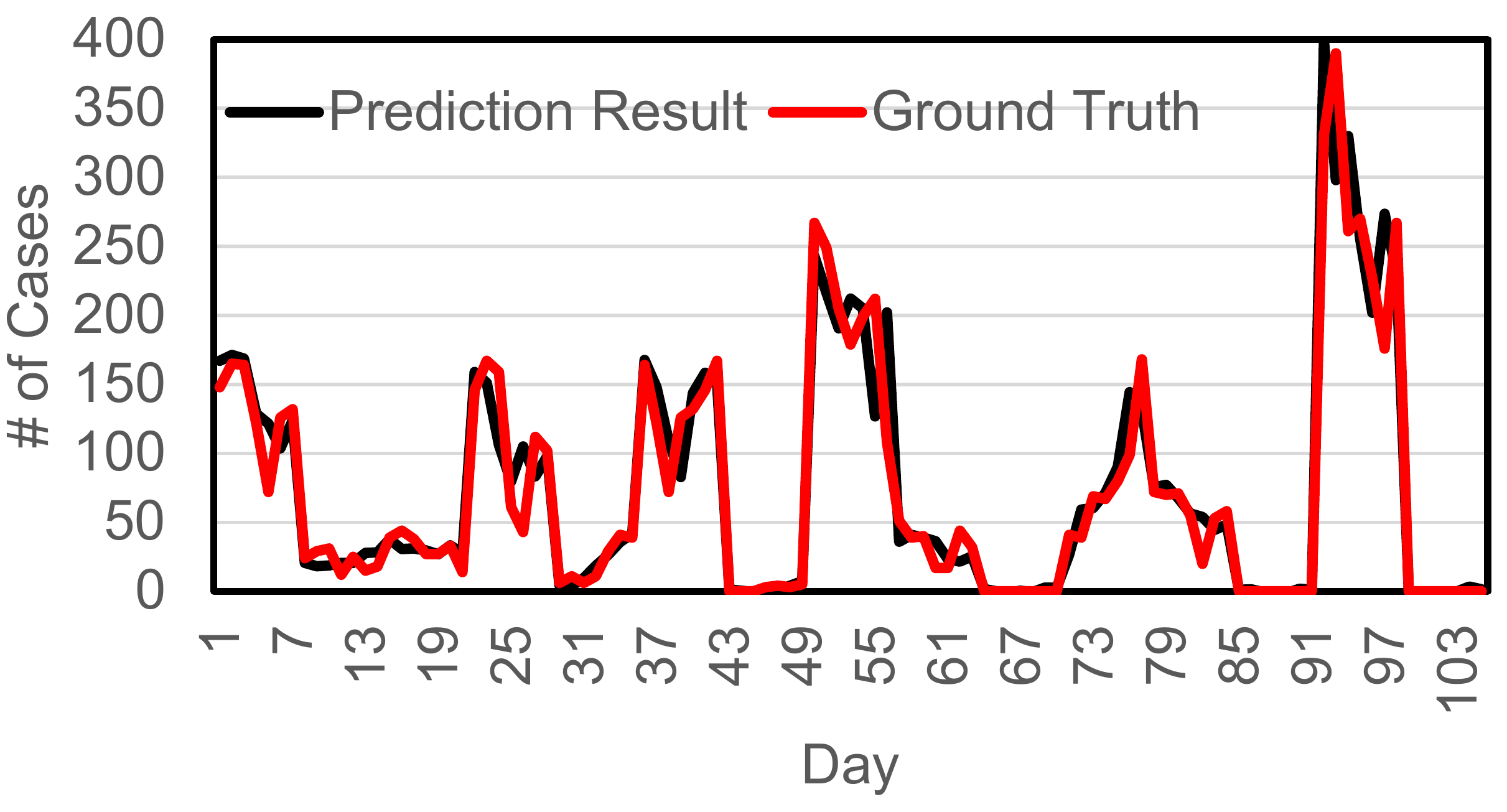}
\vspace{-0.3cm}
\caption{Prediction result and ground truth.}
\label{fig:prediction}
\vspace{-0.2cm}
\end{figure}

Fig.~\ref{fig:prediction} present the confirmed-cases prediction results, which is denoted by the black curve. In order to conduct an intuitive comparison between the prediction results with the ground truth, we also include a red curve to represent the value of ground truth. It can be seen that our prediction result is very close to the ground truth, which motivates us to get the statistic evaluation results.

\subsection{Evaluation Results and Observations}
Next, we present the key prediction quality measures for the six experiment groups (Figure~\ref{fig:ExperimentResultsFinal}). 
Note that among the three evaluation metrics, \ie $R^2$, MAE, and RMSE, 
$R^2$ is a more intuitive and objective performance indicator of the fitting effect in the regression problem. Therefore, we treat $R^2$ as our primary evaluation metric. Finally, we make several interesting observations as follows:

\begin{figure}[h]
\centering
\includegraphics[scale=0.32]{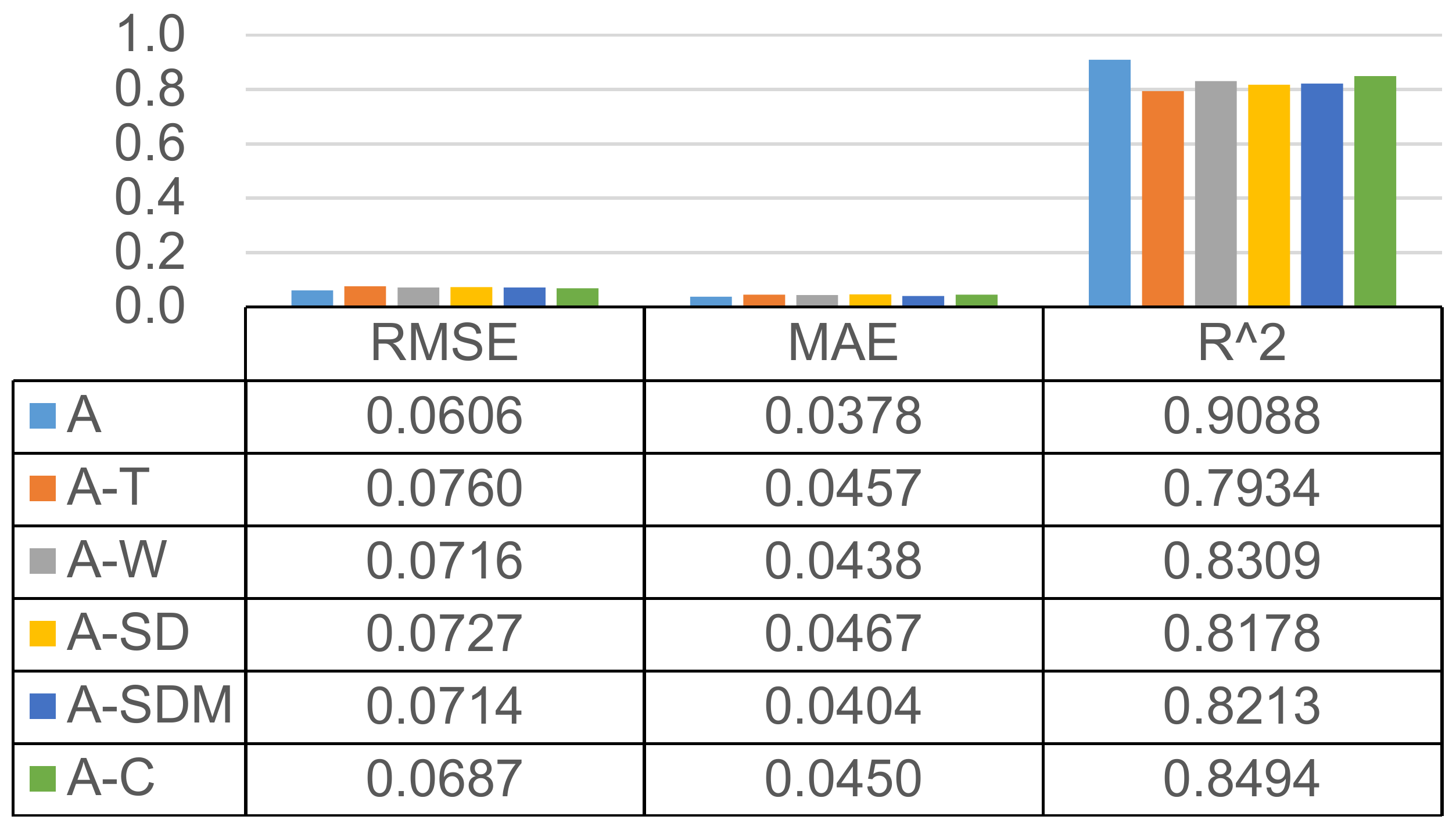}
\caption{Model prediction quality with six experiment groups of A (all selected features), A-T (without traffic volume related features), A-W (without weather-related features), A-SD (without social distancing related index on the Wayne County level), A-SDM (without social distancing related index on the Detroit level), and A-C (without crash-related features).}
\label{fig:ExperimentResultsFinal}
\vspace{-0.4cm}
\end{figure}



\noindent \textbf{\\1. } We observe that A group performs the best across all experiment groups \ie achieving highest (around 0.91) $R^2$ score and lowest MAE and RMSE. This observation verifies our hypothesis that $i)$ traffic volume data, $ii)$ weather features, $iii)$ social distancing-related data, and $iv)$ crash information are both useful and helpful for improving the effectiveness of confirmed-cases prediction.

\noindent \textbf{\\2. } Considering the difference of the $R^2$ score between A group and other five groups, we observe that: $(1)$ A-T group achieves the lowest highest score, \ie there is the biggest effectiveness difference between A group and A-T group, which proves that deleting traffic volume data could result in the most significant adverse effect on the confirmed-cases prediction, \ie traffic volume data is critical for the improvement of confirmed-cases prediction.

\begin{figure}[h]
\centering
\includegraphics[scale=0.47]{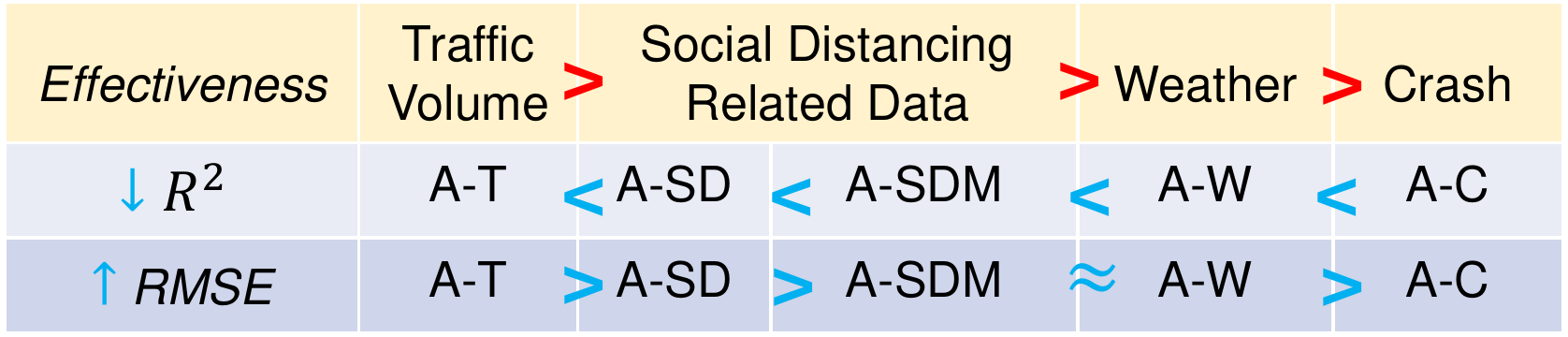}
\vspace{-0.3cm}
\caption{Effectiveness comparison between four types of data in terms of prediction improvement. The arrow indicates the higher or lower the value of the evaluation metric, the better the prediction performance.}
\label{fig:GroupEffect}
\end{figure}

\noindent \textbf{\\3. } Similar to the above observation, we get the conclusion on the effectiveness comparison between traffic volume data, social distancing related data, weather data, and crash data in terms of prediction improvement (shown in Fig.~\ref{fig:GroupEffect})---adding these four types of data can both improve the prediction performance, and traffic volume data is more effective compared with social distancing related data, then followed by weather features, and crash data seems has least impacts on the prediction performance.


\begin{tcolorbox}[left = 0mm, right = 0mm, top = 0mm, bottom = 0mm, boxsep = 1mm, arc = 0mm]
 \textbf{Observations 6:} Besides the daily case data and social distancing index, the data related to traffic volume, crashes, and weather can both be a good indicator for COVID-19 confirmed-cases prediction. The traffic volume data is very useful information regarding the prediction model.
\label{observations1}
\end{tcolorbox}

\section{Conclusion}
\label{sec:conclusion}
In this work, we collected and analyzed five types of data sets including: traffic volume, daily cases, weather information, crash features, and social distancing related data. Important observations with supporting evidence and analysis are presented to provide practical implications for authorities and decision-makers on taking preventive actions for Detroit. In terms of crashes there was a clear change in crash percentage distribution by type. During COVID-19 there was a significant increase in angle crashes which are typically more severe and indicative of more severe driver-behavior-related issues. In terms of correlations, daily cases \ie daily confirmed cases and daily death is highly related, with: $1-$ the number of transportation volume, especially for the cars, $2-$ total crashes, $3-$ social distancing index at the Wayne County level and Michigan level, and $4-$ the average temperature.
Additionally, we have trained an accuracy deep learning model, which shows the effectiveness of predicting COVID-19 confirmed-cases for the next week, \ie 7 days, with $R^2$ up to approximately 0.91. The prediction quality is tested on six experiment groups, and the prediction results also proved that adding traffic volume data, social distancing related metrics, weather information, and crash feature could both improve the prediction performance.



\bibliographystyle{IEEEtran}

\end{document}